\documentclass{article}

\evensidemargin=\oddsidemargin
\def\Title#1#2#3{%
    \baselineskip=18pt
    \begin{center}
          {\large\bf{#1} \\ }
          \bigskip\bigskip
          {#2} \\
          {#3} \\
    \end{center}}
\long\def\Abstract#1{%
         \bigskip
         \parbox{0.93\textwidth}{%
                 \begin{center}
                       {\bf Abstract} \\
                 \end{center}
                 \medskip{\baselineskip=14pt #1}
                 \vss}
         \bigskip}

\makeatletter
\renewcommand{\section}%
 {\@startsection{section}{1}{0pt}%
  {-3.25ex plus -1ex minus -.2ex}{1.5ex plus .2ex}%
  {\vspace*{5mm}\raggedright\large\bf }}
\renewcommand{\subsection}%
 {\@startsection{subsection}{2}{0pt}%
  {-2.25ex plus -.5ex minus -.2ex}{-1.5ex plus -.2ex}{\bf }}
\renewcommand{\subsubsection}%
 {\@startsection{subsubsection}{3}{0pt}%
  {-1.25ex plus -.2ex minus -.1ex}{-1.2ex plus -.2ex}{\bf }}

\makeatother

\begin{document}

\Title{Is the Copenhagen interpretation inapplicable to Quantum Cosmology?}%
{T. P. Shestakova}%
{Department of Theoretical and Computational Physics,
Southern Federal University,\\
Sorge St. 5, Rostov-on-Don 344090, Russia \\
E-mail: {\tt shestakova@sfedu.ru}}

\Abstract{It is generally accepted that the Copenhagen interpretation is inapplicable to quantum cosmology, by contrast with the many worlds interpretation. I shall demonstrate that the two basic principles of the Copenhagen interpretation, the principle of wholeness and the principle of complementarity, do make sense in quantum gravity, since we can judge about quantum gravitational processes in the Very Early Universe by their vestiges in our macroscopic Universe. I shall present the extended phase space approach to quantum gravity and show that it can be interpreted in the spirit of the Everett's ``relative states'' formulation, while there is no contradiction between the ``relative states'' formulation and the mentioned basic principles of the Copenhagen interpretation.}

\section{Introduction}
There exists an opinion that the Copenhagen interpretation of quantum theory is inapplicable to quantum cosmology, and the many worlds interpretation would be probably the only interpretation which is appropriate for this case. Some authors believe that the reason why the Copenhagen interpretation cannot be applied to quantum cosmology is the wave function reduction which they consider to be one of main ideas of this interpretation. For example, in \cite{BT} one can read:
\begin{quote}
``...the wave function collapse postulated by the Copenhagen interpretation is dynamically ridiculous, and this interpretation is difficult if not impossible to apply in quantum cosmology.''
\end{quote}
Another example of an absolutely negative attitude to the Copenhagen interpretation one can find in \cite{PNF}:
\begin{quote}
``...the Copenhagen interpretation must assume that there is a fundamental process in a measurement which must occur outside the quantum world, by an external agent in a classical domain. Of course, if we want to quantize the whole Universe, there is no place for a classical domain or any external agent outside it, and the Copenhagen interpretation cannot be applied...
Hence, if someone insists on the Copenhagen interpretation, at least in its present form, she or he must assume that quantum theory is not universal, that quantum cosmology does not make any sense at all and that we are stuck.''
\end{quote}

Of course, the answer to the question about applicability of the Copenhagen interpretation to quantum cosmology depends on what concepts one thinks to be the essence of this interpretation. I believe that the idea about the role of the observer with his instruments in quantum theory is such a concept. I would like to note that if one considers the whole Universe as a quantum object at any stage of its evolution (and not only at the very early stage) one should bear in mind a macroscopic character of the Universe, so that though there is no place for a classical domain outside the Universe, it comes into being inside it, together with observers who are willing to study it.

Speaking about the observer in the context of quantum cosmology some authors could come to misunderstandings. Indeed, we would hardly suppose that some observer could have been located in a domain of the Plank size in the Very Early Universe. This would lead to the conclusion that there are no observers inside the Universe as well as outside it (I mean the observers with their macroscopic devices, in accordance with the Copenhagen interpretation).

To clarify this point, let us start from another question: What can we expect from quantum theory of gravity? In any case, we seek for a theory that could add something to our understanding of the Universe. I think, we can expect explanation of quantum gravitational phenomena, not only those that took place in the Very Early Universe, but also in the neighborhood of a black hole and, actually, at any spacetime point. However, we can make some conclusions about these phenomena only by their macroscopic consequences that could be observed by means of our macroscopic instruments. In the Very Early Universe quantum gravitational effects started processes consequences of which we can observe billions of years later.

The only observers we can speak about staying on a solid scientific basis are human beings. The conditions, in which we, as human beings, exist, determine that we {\it have to} use macroscopic instruments. This circumstance has been taken into account yet by the founders of the Copenhagen interpretation. The situation in quantum cosmology is not trivial and very different from ordinary quantum mechanics, it requires a special treatment, but it is not obvious in advance that the Copenhagen interpretation cannot be useful in this situation.  The question about the role of the observer does make sense in quantum cosmology, just as in ordinary quantum mechanics.

In the end of 1990s the extended phase space approach to quantization of gravity \cite{SSV1,SSV2,SSV3,SSV4} was proposed. Its aim was to take into account one of the main features of gravity, namely, that most of gravitating systems do not possess asymptotic states, in contrast to systems considered in ordinary quantum field theory. Mathematically it means that asymptotic boundary conditions cannot be used in the path integral formulation of quantum gravity and ensure its gauge invariance. As a result, a wave function will depend not only on spacetime geometry, like in the Wheeler -- DeWitt quantum geometrodynamics, but also on a reference frame with respect to which this geometry is studied. Though this outcome is just a strict mathematical consequence of the mentioned feature of gravity, it may seem to be rather strange for those who believe that quantum gravity must be a gauge invariant theory not depending on the choice of a reference frame and the state of the observer. So, this result requires its assessment. In this very case the Copenhagen interpretation can help to understand the situation.

It would be wrong to say that the applicability of the Copenhagen interpretation was one of prerequisites of the extended phase space approach. Only physical features of gravity determined its essence. The interpretation comes at the last stage helping to judge the obtained results.

It would be also wrong to say that the Copenhagen interpretation is applicable to quantum cosmology just in the framework of the extended phase space approach. As I have already mentioned, the interpretation emphasizes the role of the observer as a connecting link between the theory and observations. Quantum gravity is thought to be a theory unifying general relativity and quantum theory, and in the both the observer plays a significant role. Do we have grounds to expect that results of this unification will not depend on conditions of physical experiments? We hope that quantum gravity will be a verifiable theory. But what could we observe to verify it? It must be macroscopic consequences of quantum gravitational phenomena fixed by our macroscopic instruments. Then, the main ideas of the Copenhagen interpretation may be relevant, though their generalization and future development may be required.

In the book \cite{Sud} nine interpretations of quantum mechanics are discussed, and this list is not exhaustive. It is also worth mentioning alternative formulations of quantum theory, such as the de Broglie -- Bohm theory or the decoherent (consistent) histories approach. (For applications of these approaches to quantum cosmology see, for example, \cite{PNF,Hall1}.) Mathematical content of these theories is not equivalent to the generally accepted Schr\"odinger -- Heisenberg formulation. For example, the so-called guiding equation, being an inherent component of the de Broglie -- Bohm theory, is not included into the axioms of the standard formulation of quantum mechanics. In this paper I shall restrict myself to the standard formulation without any additional postulates.

Thus, I intend to demonstrate that the Copenhagen interpretation can be applied not only to a limited class of laboratory experiments but it can be indeed useful in our search for consistent quantum theory of gravity. Moreover, I intend to discuss whether the Copenhagen interpretation is incompatible with the ``relative states'' formulation of Everett and if the choice of quantization scheme could give arguments in favor of one or another interpretation.

Starting from general points, in Sec. 2 I analyze the main ideas on which the Copenhagen interpretation is grounded, and in Sec. 3 the attention is paid to the many worlds interpretation as an alternative to the Copenhagen one. In Sec. 4 I consider some approaches to the problem of measurement and their connection to the choice of interpretation. Then I shall turn to quantum gravity, and in Sec. 5 I discuss what I call ``destruction of spacetime in quantum gravity''. I mean that the spacetime structure determined by the lapse and shift functions plays no role in the Wheeler -- DeWitt theory. It seems to be natural from the point of view of the requirement of its gauge invariance, but leads to the question: is the theory we try to quantize general relativity? Sec. 6 is entirely devoted to the extended phase space approach to quantization of gravity and its interpretation.

\section{The main ideas of the Copenhagen interpretation}
It is easy to notice that various authors treat main ideas of the Copenhagen interpretation in different ways, in a certain sense we deal with ``an interpretation of the interpretation''. So, Mukhanov \cite{Mukh} believes that the Copenhagen interpretation is based on the two points:
\begin{enumerate}
\item The idea of probability reduction in measurements.
\item The wave vector is {\it a mathematical tool} for the description of the results of measurements and has no meaning independent of concrete experiments.
\end{enumerate}
Kiefer \cite{Kiefer1} consider the following ingredients as being characteristic for the Copenhagen interpretation:
\begin{enumerate}
\item Indispensability of classical concepts for the description of the measurement process.
\item Complementarity between particles and waves.
\item Reduction of the wave packet as a formal rule without dynamical significance.
\end{enumerate}
Let us remember, however, what Bohr wrote in his ``Discussion with Einstein on epistemological problems in atomic physics'' \cite{Bohr1}:
\begin{quote}
``\ldots This crucial point\ldots\ implies the {\it impossibility of any sharp separation between the behaviour of atomic objects and the interaction with the measuring instruments which serve to define the conditions under which the phenomena appear}. In fact, the individuality of the typical quantum effects finds its proper expression in the circumstance that any attempt of subdividing the phenomena will demand a change in the experimental arrangement introducing new possibilities of interaction between objects and measuring instruments which in principle cannot be controlled. Consequently, evidence obtained under different experimental conditions cannot be comprehended within a single picture, but must be regarded as {\it complementary} in the sense that only the totality of the phenomena exhausts the possible information about the objects.''
\end{quote}
In my opinion, the Bohr's passage refers to two fundamental principles of the Copenhagen interpretation, namely, {\it the principle of wholeness (indivisibility)} of quantum phenomena and {\it the principle of complementarity}. The principle of wholeness states the impossibility to separate any quantum object from a measuring device, the object and the device must be studied together as a holistic quantum phenomenon. In his other paper \cite{Bohr2} Bohr also emphasized the feature of wholeness of a quantum phenomenon expressed in the circumstance that any attempt of its subdivision requires a modification in the experimental conditions incompatible with the definition of the phenomena under investigation.

On the other hand, physical properties of the quantum object, manifested by its interaction with measuring instruments of some type, should be supplemented by those of its properties that can be detected in experiments of another type. Understanding of the behaviour of the quantum object in various experimental conditions will give us a complete knowledge about the object as far as possible in quantum theory. Physical pictures one can observe in different experimental condition complement each other. This is an essence of the principle of complementarity. Complementarity is understood here in a broader sense then just complementarity between particles and waves.

The statements emphasized by other authors can be considered as consequences of these two principles. Indeed, as a result of measurement a new holistic system consisting of a quantum object and a measuring instrument will be formed, and a state that the quantum object had before the measurement will completely change. The existing mathematical tools of quantum mechanics do not allow us to trace the changes in details and consider the measurement as a process developing in time; we can only establish the fact of irreversible changes of the state that is expressed in the concept of ``the reduction of the state vector''. It also follows that the state vector does not make sense regardless of the specific experiments, because without experiments one cannot judge about the state of a quantum object as an element of physical reality.

In spite of many authors consider the reduction of the state vector as a basic ingredient of the Copenhagen interpretation, I would like to ask: is not the reduction a consequence of the mathematical apparatus of quantum mechanics, more exactly, is not it a consequence of the fact that the mathematical apparatus does not contain any description of a measurement process?

In his ``Principles of Quantum Mechanics'' \cite{Dirac} Dirac explains the principle of superposition by considering a superposition of two states which can be distinguished by outcomes of an observation made on some quantum system. Indeed, one cannot determine a state of a quantum system by any other way than by making observations on the system. If we have the superposition of two states, $A$ and $B$, such that some observation which is certain to lead to the results $a$ and $b$ when made on the system in the states $A$ and $B$, correspondingly, the result of the observation made on the system in the superposed state will never be different from $a$ and $b$. In this consideration the superposition principle implies the notorious ``quantum jump'' from the superposed state to one of states ($A$ or $B$) in the superposition. Let me remind that the superposition principle is one of the axioms which quantum mechanics as a mathematically consistent theory is grounded on, independently on an interpretation. From this point of view, the problem of the reduction should be understood in a broader sense as the problem of measurement, which is not a feature of the Copenhagen interpretation, but manifests itself in any other interpretation. So, in the many worlds interpretation, the main alternative to the Copenhagen one, the reduction is replaced by ``branching'', or ``splitting'', of the wave function, which is as incomprehensible as the reduction.

In \cite{Mukh} Mukhanov explains the inapplicability of the Copenhagen interpretation to quantum cosmology in the following words:
\begin{quote}
``Is {\it quantization} of the entire Universe meaningful from the point of view of the Bohr's concept of quantum theory? The answer to this question is negative. Indeed, in this case the [state] vector has no independent meaning in the absence of observer. Hence, quantization of the Universe\ldots\ is also meaningless\ldots\ because {\it outside} the Universe there are no observers and apparatus that could make measurements\ldots\ ''
\end{quote}
Everett \cite{Everett} was the first who applied quantum theory to isolated composite systems without external observers. Nevertheless, does it mean that we have to reject the Copenhagen interpretation?

\section{Everett's ``Relative state'' formulation and the many worlds interpretation}
In his seminal paper \cite{Everett} Everett wrote that his aim was
\begin{quote}
``\ldots not to deny or contradict the conventional formulation of quantum theory\ldots\ but rather to supply a new, more general and complete formulation, from which the conventional interpretation can be {\it deduced}.''
\end{quote}

Everett investigated a composite system $S$ which is composed of two subsystems $S_1$ and $S_2$. The Hilbert space ${\cal H}$ for the system $S$ is the tensor product of the Hilbert spaces ${\cal H}_1$ and ${\cal H}_2$ for the subsystems $S_1$ and $S_2$, respectively, ${\cal H}={\cal H}_1\bigotimes {\cal H}_2$. Let the sets $\{\xi_i^{S_1}\}$ and $\{\eta_j^{S_2}\}$ are complete orthonormal sets of states in the Hilbert spaces ${\cal H}_1$ and ${\cal H}_2$, then the general state of the system $S$ can be written as
\begin{equation}
\label{psi_S}
\psi^S=\sum_{i,j} a_{ij}\xi_i^{S_1}\eta_j^{S_2}.
\end{equation}
For any given state in one subsystem one can uniquely assign a {\it relative state} in the other subsystem. So, for some state $\xi_k^{S_1}$ of the subsystem $S_1$ one can define the relative state of the subsystem $S_2$ that will be
\begin{equation}
\label{rel_state}
\psi\left(S_2; {\rm rel}\;\xi_k^{S_1}\right)=N_k\sum_j a_{kj}\eta_j^{S_2},
\end{equation}
$N_k$ is a normalization constant. For each state $\xi_k^{S_1}$ the relative state (\ref{rel_state}) does not depend on the rest vectors of the orthogonal set $\{\xi_i^{S_1}\}$ and is determined uniquely by $\xi_k^{S_1}$. The state vector (\ref{rel_state}) gives the conditional probability distributions for results of all measurements in $S_2$ provided that the subsystem $S_1$ has been found to be in the state $\xi_k^{S_1}$. Using (\ref{rel_state}), the superposition (\ref{psi_S}) can be rewritten in the form:
\begin{equation}
\label{psi_S_rel}
\psi^S=\sum_i\frac1{N_i} \xi_i^{S_1}\psi\left(S_2; {\rm rel}\;\xi_i^{S_1}\right).
\end{equation}

As we can see, Everett did not use any new mathematics. One needs just several simple formulas to introduce the concept of relative states. One of the subsystems may represent a macroscopic measuring device. To illustrate this concept Everett used a toy model proposed by Neumann \cite{Neumann}. The main idea is that there exists a relation between parameters of a quantum object (in the toy  model it is the position of a particle, $q$) and parameters of a measuring instrument (in the toy model it is the position of a pointer, $r$). In the case when the pointer coordinate $r$ is determined, the wave function of the total system is given by the last equation on p.~456 in \cite{Everett}:
\begin{equation}
\label{toymod}
\psi_T^{S+A}=\int\frac1{N_{r'}}\;\xi^{r'}\!(q)\;\delta(r-r')\;dr'.
\end{equation}
This equation is an analogue of (\ref{psi_S_rel}). Here ${\xi^{r'}\!(q)}$ can be called the relative states for the states of the measuring instrument $\delta(r-r')$ when the pointer position has a concrete value $r=r'$ and $N_{r'}$ is again a normalization constant.

In fact, the concept of relative states is just a result of application of mathematical apparatus of conventional quantum theory to composite systems. In accordance with the quoted words by Everett it does not contradict the conventional formulation. I would like to emphasize that it is not interpretation, since the mathematics can be interpreted in different ways.

There is nothing in this consideration that contradicts the Copenhagen interpretation. Any changes in the experimental arrangement imply also changes in the type of interaction between the quantum object and the measuring device; mathematically it results in another basis in Hilbert subspace of the device and, accordingly, in another set of relative states of the object. The both (the object and the measuring instrument) are treated quantum mechanically but the instrument is thought to be  macroscopic.

It is the next part of the paper \cite{Everett} that gives rise to the many worlds interpretation, I mean the claim of Everett that
\begin{quote}
``\ldots with each succeeding observation (or interaction), the observer state `branches' into a number of different states.''
\end{quote}
This statement was strengthened by DeWitt who did much for the recognition of the many worlds interpretation. In \cite{DeWitt1} he wrote:
\begin{quote}
``Our universe must be viewed as constantly splitting into a stupendous number of branches, all resulting from the measurementlike interactions
between its myriads of components. Because there exists neither a mechanism within the framework of the formalism nor, by definition, an entity
outside of the universe that can designate which branch of the grand superposition is the `real' world, all branches must be regarded as equally real.''
\end{quote}
The quintessence of the many worlds interpretation can be found in \cite{Mukh}:
\begin{quote}
``Suppose, however, that {\it there exist many universes} and that the superposition describes simultaneously the results of experiments carried out in {\it different} universes, and to each concrete universe and to the result of a measurement in it there corresponds one of the terms in the superposition\ldots\ The existence of other universes is just what accounts for interference effects. Using this approach, one can interpret quantum mechanics as a theory the very {\it existence} of which is due to the existence of many worlds.''
\end{quote}

Thus, DeWitt and later Mukhanov supposed the reality of all branches and, therefore, the reality of the splitting phenomenon. At the same time, the observer does not recognize the splitting phenomenon, he stays at his own branch without having even a theoretical possibility to check the reality of other branches. If one can hope that the wave function reduction might be, in principle, described by a future theory of quantum measurement, in case of the splitting phenomenon one deals with an unverifiable hypothesis, so that this phenomenon cannot be attributed to physical reality. At the best, the hypothesis is fascinating from a philosophical point of view, but do we need to believe that all these ``myriads of universes'' are real?

As was pointed out in \cite{Grib}, the proponents as well as the opponents of the Everett interpretation are often incorrect when interpreting the splitting phenomenon, because what is implied is not the splitting of universes themselves but rather the splitting of states of universes. In this connection, let me mention the reintepretation of the Everett interpretation proposed by Mensky \cite{Mensky1,Mensky2}. As Mensky emphasized, if a superposition of quantum states exists at some time, it will be further retained, since the evolution of the system is described by a linear Schr\"odinger equation, therefore, the reduction should be excluded. But it is necessary to explain why the observer can see a single result of a measurement corresponding to just a single component of the superposition. To overcome this contradiction, Mensky appeals to the observer's consciousness. He suggested that all superposition components exist and describe different alternatives, while it is the consciousness that separates the alternatives. If a person observes one of the alternatives, he cannot see the other ones at the same time.

Here, as the author of the hypothesis of separation of the alternatives recognizes himself, we go beyond physics. Discussing the problem of measurement physicists quite often turn to the observer's consciousness (see, for example, \cite{Grib}). However, from a purely physical point of view one would come to the same effect that predicted by the concept of reduction: the observer will recognize only one of the alternative
classical pictures of the world. Therefore, the involvement of consciousness cannot help to improve our understanding what happens at a physical level when the act of measurement takes place.

One can see that the famous Everett's paper \cite{Everett} contains two different concepts: the concept of relative states, which does not give rise to doubts, and the many worlds interpretation, that can hardly help us to understand the physical phenomenon of measurement. While the many worlds interpretation is widely known and, in some sense, has overshadowed the concept of relative states, the latter one plays no role in quantum cosmology though the original Everett's motivation was to reformulate quantum theory in a form that can be applied to general relativity, in particular, to the Universe as a closed system.

\section{The problem of measurement and interpretation}
There exist various approaches to the problem of measurement. We do not know even if the measurement is a spontaneous quantum ``jump'' or a process that takes some time. The promising candidate to explain the measurement process is believed to be decoherence \cite{Mensky1,Mensky2,Zeh1,Zurek1,Zurek2}. It is based on the undoubtable fact that macroscopic systems are never isolated from their environments. It implies quantum entanglement between an apparatus and environment that leads to diagonalization of a density matrix. So, decoherence claims to explanation why one cannot observe superposed states of macroscopic objects. To explain it, orthogonality of environmental states, $\langle{\cal E}_i|{\cal E}_j\rangle=\delta_{ij}$, is required (at least approximately). In the literature one can read that ``under realistic conditions, different environmental states are orthogonal to each other'' \cite{Kiefer1}.

Some idea why the environmental states must be orthogonal could be found, for example, in \cite{Mensky1}. The environment is considered as being composed of a large number of subsystems (degrees of freedom) $\alpha$, $\beta$, \ldots, $\omega$, so that an environmental state is a product of states of subsystems, $|{\cal E}\rangle=|\alpha\rangle|\beta\rangle\ldots|\omega\rangle$. Then $\langle{\cal E}_i|{\cal E}_j\rangle$ is the product of quantities $\langle\alpha_i|\alpha_j\rangle$, $\langle\beta_i|\beta_j\rangle$, etc, each less than one in magnitude, and
$\langle{\cal E}_i|{\cal E}_j\rangle$ tends to zero.

More sophisticated models of decoherence take into account effects of scattering of each separate particle of the environment on the apparatus neglecting interactions of the environmental particles with each other and with the quantum object. Again, in this case
$\langle{\cal E}_i|{\cal E}_j\rangle$ can be presented as a product of many factors of magnitude less then one \cite{JZ} (see also \cite{Adler}).

This consideration is relied upon the assumption of quasi-independence of the subsystems when each of them can be described by its own state vector $|\alpha\rangle$, $|\beta\rangle$, etc. It seems to be in contradiction with the wholeness principle and the very idea of entanglement among a quantum object, an apparatus and environment. The well-known problem of description of quantum systems of many particles is that they are in strongly entangled, not factorizing states. As Zurek mentioned \cite{Zurek2}, not all aspects of decoherence are completely clear. Then the question remains, can more complicated models weaken the main conclusions obtained in the framework of this approach?

The second claim of the decoherence approach is that the diagonalization of a density matrix occurs without having to appeal to anything beyond
the unitary Schrodinger evolution \cite{Zurek2}. Under unitary evolution any state of the form
$(a_1|1\rangle+a_2|2\rangle)|A_0\rangle|{\cal E}_0\rangle$ will develop into the state
$a_1|1\rangle|A_1\rangle|{\cal E}_1\rangle+a_2|2\rangle|A_2\rangle|{\cal E}_2\rangle$. So, at the fundamental level the resulting state is not a mixed but superposed state which cannot be observed. Only the density matrix with environmental degrees of freedom being traced out is (almost) diagonal. It may be enough ``for all practical purposes'', but arouses criticism (see, for example, \cite{Adler}). In this connection, Zeh \cite{Zeh2} noted,
\begin{quote}
``In unitary description\ldots\ the observation of an individual outcome after decoherence can only be justified by Everett's splitting observer, since the global superposition that includes the environment\ldots\ now consists of various dynamically autonomous world components which describe different macroscopic properties.''
\end{quote}

If the state vector evolves in an unitary way, without any ``quantum jumps'', should we agree with Zeh that only the many worlds interpretation is applicable? Seemingly, yes, if we accept that ``quantum jumps'' are essential component of the Copenhagen interpretation. However, in my opinion, the problem of reduction exists because until now we do not have a satisfactory description of the measuring process, and the mathematical apparatus of quantum mechanics does not contain such a description. But I do not consider the reduction as a central idea of Bohr's interpretation. Bohr talked much that the measuring instrument must be classical, but, undoubtedly, the interaction of a quantum object with the measuring instrument takes place at the microscopic level, so that the quantum object interacts directly with quantum constituents of the instrument while the latter ones interact with each other. From this point of view, the decoherence approach is not contradictory to the Copenhagen interpretation. Zurek \cite{Zurek2} wrote:
\begin{quote}
``Decoherence is of use within the framework of either of the two major interpretations: It can supply a definition of the branches in Everett's many-worlds interpretation, but it can also delineate the border [between quantum and classical] that is so central to Bohr's point of view.''
\end{quote}

When applying the decoherence approach to quantum cosmology, an additional question arises connected with the definition of an environment, as was noted in \cite{BK}. In contrast to any experiment in a laboratory, when one considers the whole Universe as a quantum object, there is no external environment as well as there is no external observer. So, one should separate all degrees of freedom in the Universe into ``essential'' and ``auxiliary'' that cannot be done uniquely.

Among other approaches to the problem of measurement, let us mention Penrose's hypothesis of ``the objective reduction'' \cite{Pen1,Pen2}. This approach also exploits the notion of entanglement, but this time it is the entanglement between a quantum object and gravitational field. According to Penrose, any superposition of states entangled with quantum states of gravitational field is unstable and reduces to one of the states, which originally had been in the superposition, in a time defined by the uncertainty of energy of two gravitational configurations. In fact, Penrose introduces an additional postulate that does not give a solution to the problem. Also this approach cannot give arguments for or against any interpretation of quantum theory.

The existence of various approaches and opinions witnesses that the problem of measurement can be solved only if we reach thorough understanding what the act of measurement is. To my mind, it is not possible to solve it in the framework of the known axioms of quantum mechanics, and, in any case, it cannot be cleared up by the choice of any interpretation.

\section{Destruction of spacetime in quantum gravity}
According to the idea of Wheeler \cite{Wheeler}, the wave function of the Universe is determined on superspace of all possible 3-space geometries and corresponding matter fields configurations. The wave function of the Universe is a solution to the Wheeler -- DeWitt equation, which is considered as more fundamental than the Schr\"odinger equation in most approaches to quantum gravity (see \cite{Shest1} in this connection). As a consequence, the wave function does not depend on time. This well-known problem of time was thought to be the main problem of the Wheeler -- DeWitt quantum geometrodynamics and, whereas many possible solutions of the problem were suggested, none of them can be considered as an ultimate solution (for reviews on this problem, see \cite{Isham,Kuchar,SS1,SS2}).

Some theorists believe that the absence of time in quantum gravity is quite natural. This point of view stems from the statement by DeWitt \cite{DeWitt2} that
\begin{quote}
``\ldots physical significance can be ascribed only to the intrinsic dynamics of the world''.
\end{quote}
Later, the absence of time has been explained by the following reasoning \cite{Kiefer2}:
\begin{quote}
``In classical canonical gravity, a spacetime can be represented as a `trajectory' in configuration space -- the space of all three-metrics\ldots\ Since no trajectories exist anymore in quantum theory, no spacetime exists at the most fundamental, and therefore also no time coordinates to parameterize any trajectory.''
\end{quote}
The same idea one can find in \cite{Rovelli}:
\begin{quote}
``\ldots in quantum gravity the notion of spacetime disappears in the same manner in which the notion of trajectory disappears in the quantum theory of a particle.''
\end{quote}
This idea seems to be obvious, but is this analogy between 3-space geometry and coordinates of a particle enough to discard the notion of spacetime in quantum gravity? Now there is a widely shared idea that the notion of time loses its meaning at the very beginning of the Universe as well as in any domain of Plank scale. Some scientists say that there is no spacetime at the fundamental level. Another argument for it originates from Bronstein \cite{Bron} (see also \cite{RV}) who noted yet in 1936 that a spacetime structure at Planck scale cannot be determined since any attempt to do it would disturb this structure. However, the impossibility to determine its structure does not mean that there is no spacetime structure at the fundamental level at all; it just means that we do not have data to judge about this structure. It allows one to suppose that this structure may be discrete or continuous, or admit any topology, etc. At the present level of our knowledge any statement about the spacetime structure in quantum gravity including its very existence (or nonexistence) is nothing more then a presumption.

I believe that to reject the notion of time we need something more than the mentioned analogy between 3-space geometry and coordinates of a particle. In particular, the absence of time implies that no quantum gravitational phenomenon can serve as a standard of time measuring, and there is no way to define time through other quantities, e.g. probabilities. It may be interesting to note that at the dawn of quantum mechanics there existed the opinion that time has no meaning in atomic physics. One of those who advocated this opinion was the physicist and philosopher N. Campbell who suggested that time is a purely statistical conception and gave some arguments for it, but did not proposed any consistent theory. However, in 1926 he wrote \cite{Campbell}:
\begin{quote}
``\ldots Temporal conceptions are so deeply embedded in all our thoughts and language that, if we decided to abandon them entirely in discussing atomic phenomena, we should find that we could not even state adequately the problems we desire to solve\ldots While the complete solution is lacking, it will be wiser to abandon all attempts at formal consistency, and to use temporal conceptions freely, even in circumstances in which their validity is denied, if by so doing we can hope to arrive at suggestions in what direction to seek the complete solution.''
\end{quote}
When one denies the notion of time, one also implies that causality also loses its meaning in the realm of quantum gravity, being replaced by probabilistic laws. But the relationship between quantum phenomena and causality seems to be more subtle. In the Feynman approach one approximates a path integral on tiny pieces of classical trajectories, so that it is enough to know a Lagrangian describing a system under consideration to derive all its quantum behaviour. It appears that classical laws based on causality give rise to quantum description while quantum phenomena underlie the macroscopic world. In this case, classical concepts may be more then just the way of presentation of measurements results.

The idea about the absence of time in quantum gravity is in complete concordance with the result of the Wheeler -- DeWitt quantum geometrodynamics that the wave function of the Universe is independent of time. In its turns, it leads to destruction of spacetime of general relativity in quantum gravity. The structure of spacetime is determined by the lapse and shift functions which are gauge (non-physical) degrees of freedom and believed to be irrelevant in quantum gravity. By {\it destruction of spacetime} I mean that this structure plays no role in the Wheeler -- DeWitt approach (for discussion of the structure of spacetime and its role in the theory of gravity, see also \cite{Shest2}). The wave function of the Universe is declared to be gauge-invariant and {\it not depending} on the lapse and shift functions, which fix a reference frame. On the contrary, in general relativity any solution of the Einstein equations can be obtained only {\it after fixing} some reference frame. Then we have come to the question: is the theory we try to quantize general relativity or some other theory of gravity in which gauge degrees of freedom play no role?

The generally accepted answer is yes, but time and an observer in his reference frame appear only in the classical limit of quantum gravity. However, it leads us to the problem that seems to be yet more difficult. If one deals only with 3-space geometry in the domain of quantum gravity, how could it have given rise to 4-dimensional spacetime? How had time itself appeared with its attributes such as irreversibility?

A possible answer that one can find in the literature is as following \cite{Kiefer2}:
\begin{quote}
``\ldots it is obvious that the emergence of the usual notion of spacetime within quantum cosmology needs an explanation. This is done in two steps: Firstly, a semiclassical approximation to quantum gravity must be performed\ldots This leads to the recovery of an approximate Schr\"odinger equation of non-gravitational fields with respect to the semiclassical background. Secondly, the emergence of classical properties must be explained [by decoherence]\ldots''
\end{quote}
But making use of the semiclassical approximation implies that a classical spacetime {\it has already come into being} since it is a solution of classical Einstein equations that is involved into the approximate Schr\"odinger equation as a background metric. Even more, as was mentioned above, any classical solution could be found only after some reference frame had been fixed. And the most one can do is to calculate quantum corrections which are, again, based on this classical solution. It is by no means explain how time and, therefore, spacetime with its causal structure appeared from a ``timeless Universe''. Moreover, the idea about the appearance of spacetime from a ``timeless Universe'' may be interpreted mistakenly in a sense that the ``timeless Universe'' existed ``before'' the beginning of spacetime. But the initiation of time could not be a process developing in time. Then, what was it? Should we recognise that it was a ``quantum jump'' which has been discarded by the many worlds interpretation? This is an illustration of the quoted above Campbell's words that ``we could not even state adequately the problems we desire to solve''.

Even if we cannot judge about the spacetime structure, we believe that some quantum gravitational phenomena took place in the Very Early Universe, and they take place now and here in any region of Planck scale. And these phenomena give rise to our macroscopic world which we can observe now. We believe that we shall be able to judge about them {\it only} by their macroscopic consequences, otherwise it would not make sense to speak about quantum theory of gravity. Similarly, in ordinary quantum mechanics we can judge about properties of a quantum particle {\it only} by their macroscopic effects.

The wave function of the Universe determines probability distribution for the Universe to have some 3-space geometry, similarly a wave function in ordinary quantum mechanics determines probability distribution for coordinates of a particle. While every branch of the Everett wave function theoretically carries information about a sequence of all measurements made on the system, the Wheeler -- DeWitt wave function gives no indication of any measurements, understood in a wide sense as some kind of interactions. As I have already mentioned, the concept of relative states, which seemed to be specially invented for application to cosmology, plays no role in the Wheeler -- DeWitt quantum geometrodynamics. A solution to the Wheeler -- DeWitt equation describing our Universe is singled out by some boundary conditions which can be thought of as a fundamental law, extra to the known physical laws (see, for example, \cite{Hartle}). But quantum cosmology aims at explaining the physical behaviour of the Universe at the first moments of its existence when classical general relativity is not applicable \cite{Hawking}. I believe that to reach this goal it is not enough to get some probability distribution for 3-space geometry of the Universe.

Then, it makes sense to take a look at the extended phase space approach to quantization of gravity which is an alternative to the Wheeler -- DeWitt quantum geometrodynamics. And it is the Copenhagen interpretation that can help us to understand main results obtained in the framework of this approach.

\section{The extended phase space approach to quantum gravity}
In the papers \cite{SSV1,SSV2,SSV3,SSV4} a new approach to quantum geometrodynamics, the so-called ``extended phase space'' approach was proposed. Though the questions of interpretation were touched upon in \cite{SSV3}, those were not the questions of interpretation which inspired a search for a new approach. This search was motivated by the well-known problems of the Wheeler -- DeWitt quantum geometrodynamics such as the problem of time.

The Wheeler -- DeWitt quantum geometrodynamics \cite{DeWitt2} is based on three cornerstones: the Dirac approach to quantization of systems with constraints \cite{Dirac1,Dirac2}, the Arnowitt -- Deser -- Misner (ADM) parametrization \cite{ADM} and the ideas of Wheeler concerning a wave functional describing a state of gravitational field \cite{Wheeler}. Let us comment on each of these points.

The Dirac approach was the first attempt to construct a quantum theory of constrained systems. Dirac believed that it would be important to put any theory in the Hamiltonian form before quantizing it. He divided all variables into physical and nonphysical (gauge) ones, and only physical variables were included into phase space. Also he should have decided what must be the role of constraints after quantization.

The central part in this approach is given to the conjecture according to which each constraint after quantization becomes a condition on a state vector. The role prescribed to the constraints could be explained by the fact that at the classical level the constraints express gauge invariance of the theory. It was initially believed that imposing constraints at the quantum level would also ensure gauge invariance of the wave functional. However, this issue has not been thoroughly investigated and {\it gauge invariance of the theory has not been proved}.

Dirac illustrated his approach taking electrodynamics as an example. But quantum electrodynamics as a very successful and experimentally verified theory was created fully independently from the Hamiltonian formulation proposed by Dirac and was based mainly on the Lagrangian formalism and perturbation theory. Successful gauge theories of modern physics are also grounded on other theoretical and mathematical methods but not the Dirac conjecture. The predictions of these theories by no means depend on imposing constraints in an operator form as conditions on the state vector. In this sense, I would say that the Dirac approach has not been verified directly. The only theoretical result, which follows uniquely from the Dirac approach, is the Wheeler -- DeWitt quantum geometrodynamics,  but it has never been verified experimentally. Some authors argue that the Wheeler -- DeWitt equation has a correct classical limit. However, as well-known, the existing of the correct limit does not mean that the quantum theory is true, because different quantum theories may have the same classical limit. So, I consider the Dirac conjecture as {\it a postulate}, the validity of which is questionable. This point is important for the soundness of the extended phase space approach and discussed in details in \cite{Shest1,Shest3}.

In general relativity gauge variables are $g_{0\mu}$-components of metric tensor, or, in the ADM parametrization, the lapse and shift functions, $N$ and $N^i$. Gauge invariance implies that the whole theory must not depend on a choice of $N$ and $N^i$. So, a procedure of derivation of the equations for the wave function of the Universe and the wave function itself must not depend on this choice. As a matter of fact, some attempts to prove gauge invariance have been made. For example, in \cite{Barv} geometrical structure of the superspace manifold was used to inspect a possible covariant operator realization of the gravitational constraints. A metric on the superspace was defined by contracting the DeWitt supermetric with some lapse function \cite{Barv}. Meanwhile, it appeared that a desired operator realisation can be obtained only under the choice $N=1$, so that the lapse function cannot be chosen arbitrary. Whereas the resulting equations seem to be covariant, the whole procedure of their derivation is not. In a case of another choice for $N$, one would come to another form of operator gravitational constraints. Yet Hawking and Page \cite{HP} indicated that the metric on space of all three-dimensional metrics $\gamma_{ij}$ does depend on the lapse function $N$. It produces a family of the Wheeler -- DeWitt equations answering to different relations between $N$ and $\gamma_{ij}$. Hawking and Page suggested that $N$ should be considered as a field independent on $\gamma_{ij}$ that corresponds to the choice $N=1$ in \cite{Barv}.

As it is well-known, the invariance under the choice of gauge variables in the theory of gravity implies invariance under transformations of spacetime coordinates, in other words, under the choice of a reference frame. For example, if one redefines a time coordinate as $t=\varphi(\tau)$, where $\varphi(\tau)$ is an arbitrary function, it would result in the corresponding redefinition of the lapse function. On the other hand, to fix a reference frame completely one should also choose the parametrization of gauge variables, and the ADM parametrization is certainly just one of many possible parametrizations. Historically, different authors used various parametrizations of gravitational variables. One can go to some new parametrization by putting
\begin{equation}
\label{N_par}
N=v(\tilde N, \tilde N^j, g_{jk}),\quad
N^i=v^i(\tilde N, \tilde N^j, g_{jk}),\quad
\gamma_{ij}=g_{ij}.
\end{equation}
(The transition from components of metric tensor to the ADM variables belongs to transformations of this type.) From the point of view of the Dirac approach, the change of parametrization (\ref{N_par}) is not a canonical transformation, because $N$, $N^i$ are not phase space variables. However, the phase space {\it can be extended} by including into it gauge degrees of freedom on an equal footing with physical ones. At first, the idea of extended phase space was put forward by Batalin, Fradkin and Vilkovisky (BFV) \cite{BFV1,BFV2,BFV3} in the framework of path integral quantization. In the case of gauge fields one should use an effective action which includes gauge-fixing and ghost terms. The missing velocities corresponding to gauge variables can be introduced into the Lagrangian by means of special (differential) gauge conditions. It {\it actually extends} the phase space of physical degrees of freedom. Then it is possible to prove that transformations of the type (\ref{N_par}) are canonical transformations in extended phase space \cite{Shest4}.

One can look at the question of gauge invariance of the Wheeler -- DeWitt theory from another point of view. One can use the path integral approach to derive the Wheeler -- DeWitt equation. Starting points for the derivation procedure may be different. So, in the work by Barvinsky and Ponomariov \cite{BP} a path integral over reduced phase space was used, while Halliwell \cite{Hall2} followed the BFV quantization scheme. In the both papers \cite{BP,Hall2} asymptotic boundary conditions for ghosts and Lagrange multipliers of gauge fixing terms were used. The origin for the asymptotic boundary conditions is ordinary quantum field theory, in the framework of which one deals with systems with asymptotic states.  In this case, physical and non-physical degrees of freedom could be separated from each other. One cannot prove gauge invariance of the path integral and the whole theory without appealing to the boundary conditions. The Fradkin -- Vilkovisky theorem states that under the asymptotic boundary conditions the path integral is gauge independent. Based on this theorem, Halliwell in his work \cite{Hall2} was able to choose the simplest gauge condition $\dot N=0$ (the differential form of the mentioned above condition $N=1$), so that there is no explicit indication to a gauge fixing function in the equation he obtained.

It is worth noting that the theory of gravity is very different from ordinary quantum field theory. The only example of a gravitating system with asymptotic states is asymptotically flat spacetime. A universe with non-trivial topology does not possess asymptotic states, and so does a closed universe. One can conclude that one has no reason to impose the asymptotic boundary conditions in quantum gravity. This fact is crucial when deriving the Wheeler -- DeWitt equation from a path integral. In the absence of the boundary conditions gauge invariance breaks down and the quantum version of the Hamiltonian constraint loses its sense.

The ADM parametrization plays an important role because it enables one to write gravitational constraints in the form independent of gauge variables -- the lapse and shift functions. It gives rise to an illusion that the theory in which the main equations are those of constraints must not depend on a choice of gauge conditions. At the same time, the ADM parametrization introduces in 4-dimensional spacetime (3+1)-splitting that is equivalent to a choice of a reference frame, and, again, we have come to the conclusion that gauge invariance breaks down.

Concerning the original idea of Wheeler that the wave function must depend only on 3-geometry we should emphasized that in fact the wave function always depends on a concrete form of the metric.

In the situation when it has not been proved that quantum theory of gravity can be constructed in a gauge-invariant way, we cannot consider gauge degrees of freedom redundant. Then, it is preferable to use extended phase space formalism. We sought for some procedure which would enable us not to postulate, but {\it derive} the Schr\"odinger (or Wheeler -- DeWitt) equation. We have found such a procedure in the path integral approach. At this point we did have to take into account one of the main features of gravitational theory that a universe with non-trivial topology may not possess asymptotic states. Accordingly, we consider the path integral without asymptotic boundary conditions. And this is the point where the approach proposed in \cite{SSV1,SSV2,SSV3,SSV4} differs from the BFV approach. In the case when we cannot be sure that the constructed theory is gauge invariant, the Wheeler -- DeWitt equation loses its meaning as an equation that expresses gauge invariance of the theory. At the same time, independently of our notion about gauge invariance or noninvariance of the theory, the Schr\"odinger equation maintains its status as a fundamental equation of quantum theory.

I shall illustrate the extended phase space approach with a model with a finite number of degrees of freedom. Models of this type have been used by many authors, see, for example, \cite{Hall2}. The action for the model can be written in the following form:
\begin{equation}
\label{action1}
S=-\!\int\!dt\,\left[\displaystyle\frac1{2N}\gamma_{ab}\dot q^a\dot q^b-NU(q)
 +\pi\left(\dot N-\frac{\partial f}{\partial q^a}\dot q^a\right)+N\dot{\bar\theta}\dot\theta\right].
\end{equation}

\noindent Here $q=\{q^a\}$ stands for physical variables and $N$ denotes a gauge variable (it may be, for example, the lapse function), $\theta$, $\bar\theta$ are the Faddeev -- Popov ghosts. We use the equation
\begin{equation}
\label{diff_form}
\dot N-\frac{\partial f}{\partial q^a}\dot q^a=0,
\end{equation}

\noindent which is a differential form of the gauge condition
\begin{equation}
\label{gauge}
N=f(q)+k;\quad
k={\rm const},
\end{equation}

\noindent $\pi$ is a Lagrange multiplier of the gauge condition (\ref{diff_form}). To consider an arbitrary parametrization of a gauge variable we introduce the function $v(\tilde N, q)$,
\begin{equation}
\label{param}
N=v(\tilde N, q),
\end{equation}

\noindent $\tilde N$ is a new gauge variable which obeys the condition
\begin{equation}
\label{newgauge}
\tilde N=F(q)+k.
\end{equation}

Then, the action will look like
\begin{equation}
\label{action2}
S=-\!\int\!dt\,\left[\displaystyle\frac12\frac1{v(\tilde N, q)}\gamma_{ab}\dot q^a\dot q^b-v(\tilde N, q)U(q)
 +\pi\left(\dot{\tilde N}-\frac{\partial F}{\partial q^a}\dot q^a\right)+w(\tilde N, q)\dot{\bar\theta}\dot\theta\right];
\end{equation}
\begin{equation}
\label{w_deff}
w(\tilde N, q)=v(\tilde N, q)\left(\frac{\partial v}{\partial\tilde N}\right)^{-1}.
\end{equation}

Turning to the quantization procedure, we derived the Schr\"odinger equation from the path integral with the effective action (\ref{action2}) and {\it without} asymptotic boundary conditions by a standard method originated by Feynman \cite{Feynman,Cheng}. The Schr\"odinger equation can be written in the following way:
\begin{equation}
\label{SE1}
i\,\frac{\partial\Psi(\tilde N,q,\theta,\bar\theta;\,t)}{\partial t}
 =H\Psi(\tilde N,\,q,\,\theta,\,\bar\theta;\,t);
\end{equation}
\begin{equation}
\label{H}
H=-\frac1{2M}\frac{\partial}{\partial Q^{\alpha}}MG^{\alpha\beta}\frac{\partial}{\partial Q^{\beta}}
  -\frac1{w(\tilde N, q)}\frac{\partial}{\partial\theta}\frac{\partial}{\partial\bar\theta}+v(\tilde N, q)(U(q)-V[F]);
\end{equation}
\begin{equation}
\label{Q0}
Q^{\alpha}=(\tilde N, q^a),\quad
\alpha=(0, a);
\end{equation}
\begin{equation}
\label{Galphabeta}
G^{\alpha\beta}=v(\tilde N, q)\left(
\begin{array}{cc}
\displaystyle\frac{\partial F}{\partial q^a}\frac{\partial F}{\partial q_a}&\displaystyle\frac{\partial F}{\partial q^b}\\
\displaystyle\frac{\partial F}{\partial q^a}&\gamma^{ab}
\end{array}
\right);
\end{equation}

\noindent $M$ is the measure in the path integral; $V[F]$ is a quantum correction to the potential $U$ which depends on the chosen parametrization and the gauge fixing function $F(q)$, it has the same origin as the term with the scalar curvature of configurational space in \cite{Cheng}. The explicit form of $V[F]$ for an arbitrary parametrization of the gauge variable is given in \cite{Shest5}. The wave function is defined on extended configurational space with the coordinates $\tilde N,\,q^a,\,\theta,\,\bar\theta$.

The general solution to the Schr\"odinger equation (\ref{SE1}) has the structure:
\begin{equation}
\label{GS}
\Psi(\tilde N,\,q,\,\theta,\,\bar\theta;\,t)
=\int\Psi_k(q,\,t)\,\delta(\tilde N-F(q)-k)\,(\bar\theta+i\theta)\,dk.
\end{equation}

\noindent It is a superposition of eigenstates of a gauge operator,
\begin{equation}
\label{k-vector}
\left(\tilde N-F(q)\right)|k\rangle=k\,|k\rangle;\;\;
|k\rangle=\delta\left(\tilde N-F(q)-k\right).
\end{equation}

The function $\Psi_k(q,\,t)$ is a solution to the equation
\begin{equation}
\label{phys.SE}
i\,\frac{\partial\Psi_k(q;\,t)}{\partial t}
 =H_{(phys)}[F]\Psi_k(q;\,t),
\end{equation}
\begin{equation}
\label{phys.H-A}
H_{(phys)}[F]=\left.\left[-\frac1{2M}\frac{\partial}{\partial q^a}v(\tilde N, q)M\gamma^{ab}\frac{\partial}{\partial q^b}
 +v(\tilde N, q)(U(q)-V[F])\right]\right|_{\tilde N=F(Q)+k}.
\end{equation}

Let us consider the general solution (\ref{GS}) from the viewpoint of Everett's ``relative state'' formulation. We can compare the expressions (\ref{toymod}) and (\ref{GS}). In of the superposition (\ref{toymod}) each element corresponds to a state with the only degree of freedom of the measuring apparatus (the pointer position $r$) being determined. In general relativity to arrive at a conclusion about spacetime geometry the observer needs to fix a reference frame. The reference frame plays the role of a measuring device in general relativity. In the superposition (\ref{GS}) each element corresponds to a state with the only gauge degree of freedom $\tilde N$ being determined, it implies that some processes in the physical subsystem define a time scale by means of the functions $v(\tilde N, q)$, $F(q)$ (see (\ref{param}), (\ref{newgauge})). The general solution (\ref{GS}) is a packet over $k$. Eq. (\ref{phys.SE}) does not fix a form of the packet, but it should be sufficiently narrow for $\Psi_k(q,\,t)$ to be a normalizable function \cite{SSV4}. The spread of $k$ means that the reference frame cannot be fixed absolutely precisely in quantum gravity and time intervals between two hypersurfaces cannot be measured with arbitrary accuracy. Thus, the function $\Psi_k(q,\,t)$ can be thought as describing a relative state of the physical subsystem for the reference frame fixed by the condition (\ref{newgauge}).

In accordance with the idea of many worlds, each term in the superpositions (\ref{toymod}) and (\ref{GS}) corresponds to some universe, but {\it it is not required} by the concept of relative states. In (\ref{toymod}) the function ${\xi^{r'}\!(q)}$ describes a relative state of the quantum subsystem {\it under the condition} that the pointer of the measuring apparatus is in the position $r'$. Different terms in (\ref{toymod}) answer various results of measurements which could be done in different universes (if the existence of many universes corresponds to your philosophic belief), or, the measurements could be done {\it in the same Universe} in identical physical conditions. The latter does not contradict the Copenhagen interpretation. Similarly, different terms in (\ref{GS}) answer various results of observations done in identical physical conditions, namely, in the same reference frame (up to small quantum fluctuations), since we suppose that the packet over $k$ is sufficiently narrow.

If one wishes to measure another physical quantity, say, the momentum of a particle instead of its position, one would come to some other superposition in lieu of (\ref{toymod}) and another set of relative states, that answers new physical conditions and changes in interaction between the quantum subsystem and the measuring device. Likewise, the change of the reference frame (the choice of another condition from the class (\ref{newgauge})) would be equivalent to making use of another basis (\ref{k-vector}) and result in some different function $\Psi_k(q,\,t)$ describing the physical Universe from the viewpoint of the observer in a different reference frame. Then, different observers would see different physical phenomena.

This conclusion seems to be quite radical, but it is in agreement with the spirit of general relativity. It is also in correspondence with the two basic principles of the Copenhagen interpretation: the principle of complementarity, since physical pictures viewed by various observers complement each other, and the principle of wholeness, because in this approach we deal with the holistic system including the physical Universe as well as the observer studying this Universe. On the other hand, though the analogy between (\ref{toymod}) and (\ref{GS}) might not be complete, our results can be considered as an attempt to give a mathematical realization of Everett's concept of relative states.

If one agrees that measuring devices in quantum gravity relate to some reference frame, there must exist some physical realization of the reference frame, that is a physical constituent the state of which affects the rest of the Universe. From the mathematical point of view, the reference frame is described by a gauge fixing term in the action. Variation of the action leads to additional term in the Einstein equations that can be interpreted as some medium. For example, in the simplest isotropic model with the gravitational part of the action
\begin{equation}
\label{isotr.act}
S_{(isotr)}=-\!\int\!dt\,\left(\frac12\frac{a\dot a^2}N-\frac12Na\right)
\end{equation}

\noindent introducing the gauge condition for the lapse function
\begin{equation}
\label{isotr.gauge}
N=a^{n-3}
\end{equation}

\noindent results in an additional (gauge-noninvariant) term to energy-momentum tensor that corresponds to the me\-dium with the equation of state $p=\left(\displaystyle\frac n3-1\right)\varepsilon$ (see \cite{Shest6}). Different values of $n$ cover well-known cases of equations of states, in particular, $n=4$ corresponds to radiation, while $n=0$ is the case of the Universe with a non-zero cosmological constant. At the classical level a medium with the same properties can be presented by the following additional term in the action:
\begin{equation}
\label{med.act}
S_{(medium)}=-\!\int\!dt\,Na^3\frac{\varepsilon_n}{a^n};\quad
{\varepsilon_n}={\rm const}.
\end{equation}

\noindent Let us introduce, as it has been done above, the arbitrary parametrization $N=v(\tilde N, a)$. The action for the model with the additional term (\ref{med.act}) is
\begin{equation}
\label{action-3}
S=-\!\int\!dt\,\left[\frac12\frac{a\dot a^2}{v(\tilde N,\,a)}-\frac12v(\tilde N,\,a)a
 +v(\tilde N,\,a)\frac{\varepsilon_n}{a^{n-3}}\right].
\end{equation}

\noindent Then we obtain the gravitational constraint in the form
\begin{equation}
\label{gen-constr}
\frac{\partial v}{\partial\tilde N}\left(\frac1{2a}p_a^2
 +\frac12a-\frac{\varepsilon_n}{a^{n-3}}\right)=0.
\end{equation}

\noindent Of course, at the classical level constraints corresponding to different functions $v(\tilde N,\,a)$ are equivalent. We can choose
\begin{equation}
\label{param2}
v(\tilde N,\,a)=\tilde Na^{n-3},
\end{equation}

\noindent and Eq.(\ref{gen-constr}) would be
\begin{equation}
\label{constr-4}
\frac12a^{n-4}p_a^2+\frac12a^{n-2}-\varepsilon_n=0.
\end{equation}

We can also write down the Hamiltonian -- Jacobi equation,
\begin{equation}
\label{Ham-J}
\frac12a^{n-4}\left(\frac{dS}{da}\right)^2+\frac12a^{n-2}-\varepsilon_n=0.
\end{equation}

At the same time, in the case of the parametrization (\ref{param2}) and the gauge condition $\tilde N=1$, which are equivalent, taken together, to the gauge condition (\ref{isotr.gauge}) for the lapse function, the stationary Schr\"odinger equation for a physical part of the wave function reads
\begin{equation}
\label{Schro-4}
-\frac12a^{\frac n2-2}\frac d{d a}
  \left(a^{\frac n2-2}\frac{d\Psi}{d a}\right)
 +\frac12a^{n-2}\Psi=E\Psi.
\end{equation}

\noindent The Hamiltonian eigenvalue $E$ is associated with the constant $\varepsilon_n$. Restoring the Planck constant in the Schr\"o\-dinger equation and seeking for the quasiclassical solution $\Psi=A\exp\left(\displaystyle\frac{iS}{\hbar}\right)$, one would come in the zero order in $\hbar$ to the Hamiltonian -- Jacobi equation (\ref{Ham-J}) that shows that the quasiclassical limit is correct.

Thus, the pure mathematical approach predicts the existence of a medium the properties of which are determined by gauge conditions and there is some dependence between the properties of this medium and those of the rest Universe described by the physical part of the wave function. Let me remind that Landau and Lifshitz \cite{LL} described the reference frame in general relativity as an infinite number of bodies with arbitrarily running clocks fixed on them which fill all the space like some sort of ``medium''. I would like also mention the well-known fact from quantum field theory that even in Minkowski spacetime the observer in an accelerated reference frame and the observer in an inertial reference frame deal with different vacuum states. Their detectors fix different physical picture. From this point of view it is reasonable to expect that a similar situation would take place in quantum gravity. I imply that the medium discussed above phenomenologically reflects some properties of vacuum in a given reference frame. A transition to another reference frame means the change of interaction between measuring devices and a quantum Universe, that results in a different picture observed in accordance with the Copenhagen interpretation.

In conclusion, I would like to touch upon the question, can we come to the Wheeler -- DeWitt equation starting from the Schr\"odinger equation obtained above? Yes, we can do it under the following condition:
\begin{itemize}
\item one should put $E=0$ in stationary solutions to the Schr\"odinger equation for the physical part of the wave function (\ref{phys.SE}) that is equivalent to rejecting time evolution of the wave function;
\item one should choose the ADM parametrization (return to the lapse function, see (\ref{param}));
\item one should choose the gauge condition $N=1$ discussed above.
\end{itemize}
In this case Eq.(\ref{phys.SE}) would reduce to the Wheeler -- DeWitt equation
\begin{equation}
\label{WDW.eq}
H_{WDW}\Psi(q)=0,
\end{equation}

\noindent with the Hamiltonian operator
\begin{equation}
\label{WDW_Ham}
H_{WDW}=-\frac1{2M}\frac{\partial}{\partial q^a}M\gamma^{ab}\frac{\partial}{\partial q^b}+U(q).
\end{equation}

So, from the point of view of the extended space approach, the Wheeler -- DeWitt equation and its solutions answer just the particular case, namely, the particular choice of the parametrization and the gauge condition and the zero eigenvalue of the Hamiltonian. If one declare that the Wheeler -- DeWitt equation is more fundamental, it means rejecting all other cases and all other solutions to the Schr\"odinger equation (\ref{SE1}). At the present level of development of the theory, it does not seem to be grounded enough. After all, we recognize the Schr\"odinger equation as fundamental just because making use of this equation it is possible to predict the results of quantum mechanical experiments. In my opinion, in the situation with quantum gravity, when we do not have experimental or observational data at our disposal, it would be incorrect to give preference to the Wheeler -- DeWitt equation excluding all other possibilities.

\section{Conclusions}
In this paper the extended phase space approach to quantization of gravity has been briefly presented. In the framework of this approach we have come to the conclusion that the description of the Universe in quantum geometrodynamics appears to be dependent on a chosen reference frame representing the observer in the theory of gravity. This conclusion seems to contradict the generally accepted point of view that the description of the Universe must be gauge-invariant. However, even when one describes quantum fields in Minkowski spacetime, a physical picture will depend on what reference frame, inertial or accelerating, is chosen. Already in classical general relativity an observer in the neighborhood of a black hole and another one at asymptotical infinity from it will see different physical phenomena. It is natural to expect that in a future theory of quantum black holes a dependence of physical phenomena on the observer's state will also be presented. Indications to it can be found in quantum field theory in curved spacetime. Then, one can suppose that interaction between an object under investigation and the observer with his measuring devices in a given reference frame should be taken into account in the full quantum theory of gravity.

In the Wheeler -- DeWitt approach one can find no vestiges of the observer. The Wheeler -- DeWitt equation enables one to get, under some additional assumptions, a probability distribution for 3-space geometry. However, it cannot offer any reasonable explanation how our 4-dimensional spacetime could have appeared from a timeless early Universe. It remains to be a fundamental problem of modern cosmology.

Our attempt to take into account the features of gravity, namely, the absence of asymptotic states, has led to the question about the role of the reference frame in quantum theory of gravity, the question, which is closely related with the one about the influence of physical conditions on quantum measurements. So, I would say that the extended phase approach favors the Copenhagen interpretation, since it is the Copenhagen interpretation that points to the central role of the interaction between an object and a measuring instrument in our understanding of quantum phenomena. The main idea of it is the unavoidable interaction between the object and the apparatus, but not the statement about the reduction of the wave function. Though the Copenhagen interpretation cannot give us a profound comprehension of this interaction, it may prompt a direction for our reflections. In this sense I believe that the Copenhagen interpretation can be useful in our search for consistent quantum theory of gravity, in particular, in our understanding of quantum cosmology. As concerns the many worlds interpretation, its acceptance or rejection is fully determined by the creed of a physicist, his belief in existence (or nonexistence) of many universes. The interpretation of quantum mechanics as a theory based on the existence of many worlds, that cannot be validated in principle, even in future, does not answer to the strict scientific criterion of testability.

On the other hand, I have argued that Everett's concept of relative states does not contradict the Copenhagen interpretation. Moreover, in the extended phase space approach we can interpret the state of the physical Universe as a relative state for a chosen reference frame. And, though the analogy between the notion of the relative state by Everett and the description of the physical Universe in the proposed approach may not be complete, it again suggests the idea about the interrelationship between properties of the object of interest and physical conditions under which the object is studied.

\section*{Acknowledgements}
I am grateful to Claus Kiefer for discussions during my visit to the University of Cologne supported by a joint grant of the Ministry of Education and Science of Russian Federation and DAAD. Though I know that Claus Kiefer does not agree with my approach to quantization of gravity and, especially, my opinion about interpretations of quantum theory, our discussions were a starting point for writing this paper.

Some parts of this paper is based on the talk given in the University of Geneva that was supported by a grant from the John Templeton Foundation to visit the Geneva Center of the project ``Space and Time after Quantum Gravity''.

\end{document}